\pdfoutput=1

\documentclass[11pt]{article}

\usepackage[preprint]{acl}

\usepackage{times}
\usepackage{latexsym}
\usepackage[T1]{fontenc}
\usepackage[utf8]{inputenc}
\usepackage{microtype}
\usepackage{inconsolata}
\usepackage{graphicx}
\usepackage{amsmath}
\usepackage{booktabs}
\usepackage{multirow}

\makeatletter
\def\thebibliography#1{\vskip\parskip%
\vskip\baselineskip%
\def\baselinestretch{1}%
\ifx\@currsize\normalsize\@normalsize\else\@currsize\fi%
\vskip-\parskip%
\vskip-\baselineskip%
\section*{References\@mkboth
 {References}{References}}\list
 {}{\setlength{\labelwidth}{0pt}\setlength{\leftmargin}{\parindent}
 \setlength{\itemindent}{-\parindent}%
 \setlength{\itemsep}{1pt plus 1pt minus 0.5pt}%
 \setlength{\parsep}{1pt plus 0.5pt minus 0.5pt}}
 \def\newblock{\hskip .11em plus .33em minus -.07em}
 \sloppy\clubpenalty4000\widowpenalty4000
 \sfcode`\.=1000\relax}
\makeatother

\newcommand{\projectrootprefix}{}
\IfFileExists{figures/overview_figure.pdf}{}{\renewcommand{\projectrootprefix}{../}}
\let\oldincludegraphics\includegraphics
\renewcommand{\includegraphics}[2][]{\oldincludegraphics[#1]{\projectrootprefix#2}}
\let\oldinput\input
\renewcommand{\input}[1]{\oldinput{#1}}

\hypersetup{pdftitle={Learning from Convenience Samples: Fine-Tuning LLMs to Predict Voting Choices from Biased Survey Samples},
  pdfauthor={Tobias Holtdirk, Dennis Assenmacher, Arnim Bleier, Claudia Wagner}}

\title{Learning from Convenience Samples: Fine-Tuning LLMs to Predict Voting Choices from Biased Survey Samples}

\author{Tobias Holtdirk$^{1,2}$ \quad Dennis Assenmacher$^{3}$ \quad Arnim Bleier$^{3}$ \quad Claudia Wagner$^{3,4,5}$ \\[4pt]
  $^{1}$LMU Munich \quad
  $^{2}$Munich Center for Machine Learning (MCML) \\
  $^{3}$GESIS -- Leibniz Institute for the Social Sciences \\
  $^{4}$RWTH Aachen University \quad
  $^{5}$Complexity Science Hub Vienna \\[4pt]
  \texttt{tobias.holtdirk@lmu.de} \\
  \texttt{\{dennis.assenmacher,arnim.bleier,claudia.wagner\}@gesis.org}}

\begin{document}
\maketitle

\begin{abstract}
   Survey research increasingly relies on convenience samples as rising costs and nonresponse make representative probability sampling more difficult. This creates a central challenge for public opinion research and electoral studies: how to recover reliable population-level inferences from socially and demographically biased data. While recent work has explored the use of large language models (LLMs) for simulating survey responses, most studies focus on the U.S. population and assume relatively favourable data conditions with limited or random missingness.
   In this work, we investigate whether LLMs can leverage contextual knowledge about socio-demographic and political patterns to predict vote choice in the German population under sampling constraints. We systematically vary both the size and composition of training data, ranging from large representative samples to small, biased convenience samples, and compare zero-shot prompting and supervised fine-tuning against established tabular classifiers such as CatBoost.

   Our results show that when representative training data are available, fine-tuned LLMs achieve predictive performance comparable to strong tabular baselines and substantially outperform zero-shot prompting. Under small and biased convenience samples, however, fine-tuned LLMs more accurately recover both individual-level vote choice and population-level distributions than zero-shot approaches and often outperform tabular methods. These findings suggest that fine-tuned LLMs can combine sample-specific information with contextual knowledge acquired during pre-training, enabling more robust inference under conditions of sampling bias and missing data.
   
\end{abstract}

\begin{figure*}[ht]
\centering
    \includegraphics[width=0.9\textwidth]{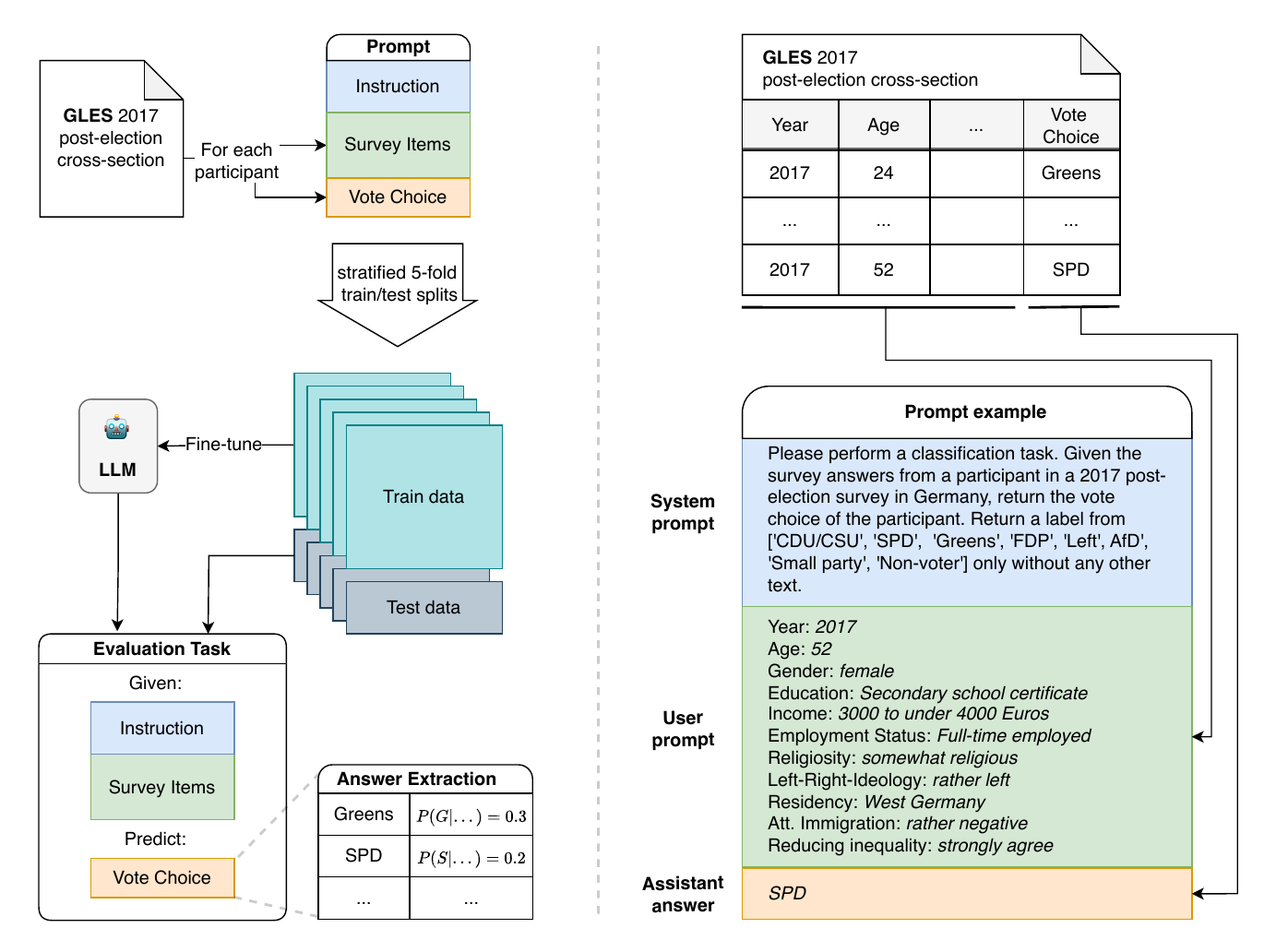}
    \caption{\textbf{Approach to simulate survey answers with LLMs.} We prompt and fine-tune LLMs to extract their answer for the question ``Which party did you vote for in the federal election?''. We compare LLM-generated answers with real survey answers. The prompt design consists of three parts: the system prompt, the user description, and the answer. For fine-tuning, we provide the real vote-choice labels of training participants as assistant completions in the depicted format. For the evaluation, we only present the system prompt and the user prompt to the model and require the model to generate the answer.
    }
    \label{fig:experiment_overview}
\end{figure*}

\section{Introduction}\label{introduction}

For decades, the empirical social sciences have relied on probability-based survey samples as one of their main data sources. However, survey-based research is witnessing transformative changes, including rising costs, declining participation, and new modes of participation \citep{berinskyMeasuringPublicOpinion2017,couperNewDevelopmentsSurvey2017}. These challenges particularly affect surveys on political opinions and voting behaviour, where systematic nonresponse biases can impair election forecasting~\citep{feldmanWhoArePeople2022,clintonReluctantRepublicansEager2022}.
As representative sampling becomes more difficult and costly, researchers increasingly turn to \emph{convenience samples}—non-probability samples in which participants are selected because they are easy to access or inexpensive to recruit.

Predicting population behaviour or responses from convenience samples can be understood as a form of distribution shift, where the observed sample on which prediction models are trained is not representative of the target population that is present in the test data. In this context, large language models (LLMs) offer an interesting alternative to traditional machine learning methods because they can draw not only on the biased training data of the convenience sample but also on patterns learned during large-scale pre-training.
In principle, exposure to broad pre-training data could compensate for subpopulations that are under-represented in the convenience sample.

Whether this potential is realised, however, depends on how LLMs are used. When simply prompted zero-shot, LLM responses can be biased and lack variation \citep{bisbeeSyntheticReplacementsHuman2024}, often displaying left-leaning tendencies in political orientation \citep{santurkarWhoseOpinionsLanguage2023,heydeVoxPopuliVox2024}. Zero-shot LLMs also face technical issues such as first-token probabilities not matching text responses \citep{wangMyAnswerFirstToken2024} and prompt sensitivity \citep{bisbeeSyntheticReplacementsHuman2024}, suggesting zero-shot prompting alone may be insufficient for survey prediction.
Advances in the efficiency of fine-tuning, such as quantisation \citep{dettmersQLoRAEfficientFinetuning2023} and parameter-efficient adaptation \citep{huLoRALowRankAdaptation2022,liuFewShotParameterEfficientFineTuning2022}, together with strong open-source models \citep{dubeyLlama3Herd2024}, make fine-tuning a more accessible alternative that can combine the pre-trained knowledge with the available sample data, addressing these issues while allowing local deployment for confidential survey data.

In this paper, we predict the voting behaviour of respondents in the German Longitudinal Election Study (GLES) \citep{glesGLESCrossSection200920172020} with LLMs (see Figure~\ref{fig:experiment_overview}) and address two research questions:
\begin{itemize}
  \item[] \textbf{RQ1:} Are LLMs more accurate and robust than tabular classifiers when \textbf{large, random samples} are available for training/fine-tuning?
  \item[] \textbf{RQ2:} Are LLMs more accurate and robust than tabular classifiers when \textbf{only small, biased convenience samples} are available for training/fine-tuning?
\end{itemize}

\begin{figure}
  \centering
  \includegraphics[width=0.99\columnwidth, trim=0 30 0 30, clip]{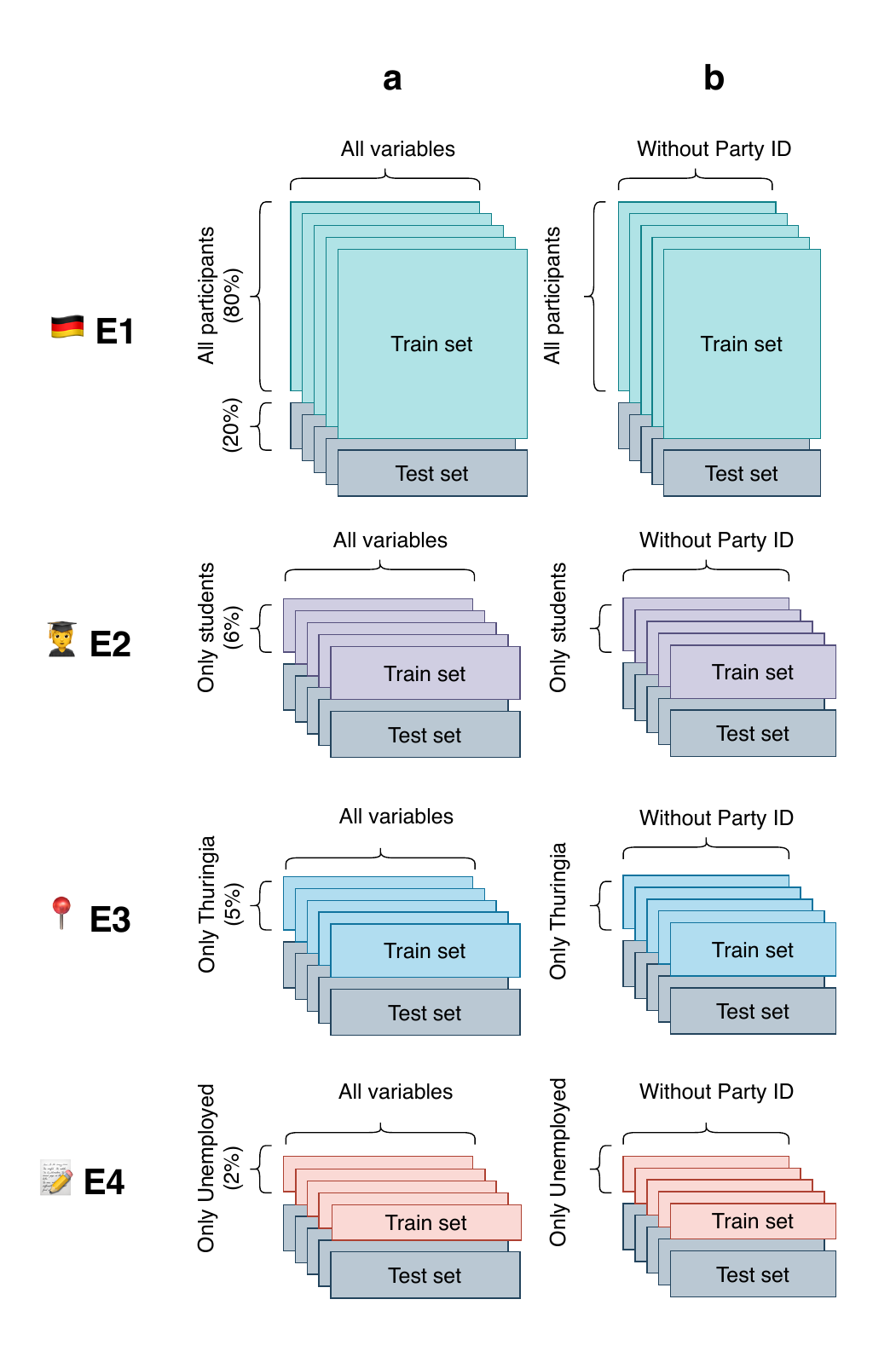}
  \caption{\textbf{Experiments.} Each experiment uses a different subset of the GLES data, reduced either by removing survey variables        (columns) or participants (rows).}
  \label{fig:data_subsets}
\end{figure}

We compare LLMs (zero-shot and fine-tuned) against tabular classifiers (random forest, CatBoost, logistic regression) across four scenarios (Figure~\ref{fig:data_subsets}). All experiments predict the vote choice of the same 20\% randomly selected GLES participants but vary the training set drawn from the remaining 80\%: \textbf{E1} uses the full 80\%; \textbf{E2} restricts training to students; \textbf{E3} to participants from Thuringia, an East German state whose 2017 voting pattern deviated markedly from the national average; \textbf{E4} to unemployed participants. E2--E4 test to what extent LLMs can learn from small, biased convenience samples and generalise to the broader population.

Our results show that when large random samples are available (E1), fine-tuned LLMs match tabular classifiers, outperform zero-shot LLMs, and remain robust when key predictors are removed. When only small, biased convenience samples are available (E2, E3), fine-tuning small open-source LLMs (3B--8B) generally yields more accurate individual-level and aggregate predictions than zero-shot LLMs or tabular baselines, because these models combine sample-specific patterns with broad contextual knowledge acquired during pre-training.

\section{Related Work}

\paragraph{Zero-shot opinion prediction.}
Most prior work uses zero-shot prompting, typically on OpenAI models. \citet{argyleOutOneMany2023} reported that GPT-3, conditioned on socio-demographic backstories from the American National Election Studies (ANES), closely reproduces vote-choice patterns in the aggregate and across most demographic subgroups, though they deliberately do not evaluate individual-level correspondence. Later work paints a less optimistic picture: \citet{santurkarWhoseOpinionsLanguage2023} showed systematic biases toward certain demographic groups and political orientations on the American Trends Panel; \citet{dominguez-olmedoQuestioningSurveyResponses2023} showed on American Community Survey (ACS) data that zero-shot survey responses are governed by ordering and labelling biases and become near-uniform once these biases are adjusted for; and \citet{bisbeeSyntheticReplacementsHuman2024} concluded that current LLMs cannot replace human ANES respondents. Outside the U.S., \citet{heydeVoxPopuliVox2024} and \citet{quPerformanceBiasesLarge2024} report lower performance for German and other non-U.S.\ populations.

\paragraph{Fine-tuning for opinion prediction.}
Several studies fine-tune or adapt open-weight language models on domain-specific data: \citet{jiangCommunityLMProbingPartisan2022} fine-tune GPT-2 on partisan Twitter communities, \citet{chuLanguageModelsTrained2023} adapt BERT to US media-outlet content, and \citet{ahnert2025extracting} fine-tune LLMs on Twitter time slices. Recent work trains on closed-ended response distributions rather than individual responses \citep{caoSpecializingLargeLanguage2025,suhLanguageModelFineTuning2025}. Our approach enables training on both closed- and open-ended questions and is compatible with fine-tuning APIs, providing a more general framework.

\paragraph{LLMs for imputation.}
Pre-trained language models have been fine-tuned for imputation \citep{meiCapturingSemanticsImputation2021}, and LLMs have been applied to data-wrangling tasks including imputation more broadly \citep{narayanCanFoundationModels2022}; the HELM benchmark \citep{liangHolisticEvaluationLanguage2023} includes a corresponding imputation scenario.
Existing work largely focuses on data missing completely at random. An exception is \citet{kimAIAugmentedSurveysLeveraging2023}, who train a dense network on LLM embeddings for U.S.\ surveys. In contrast, we directly fine-tune instruction-tuned LLMs and evaluate on German data, which is known to be more challenging \citep{heydeVoxPopuliVox2024}.

\section{Methods and Experiments}

We predict the target variable ``vote choice'' using survey variables that the literature on voting behaviour in Germany identifies as important predictors \citep{schmitt-beckNewEraElectoral2022,schoenVotersVotingContext2017,kleinGesellschaftlicheWertorientierungenWertewandel2014}; the selected items are detailed in Section~\ref{sec:method_data}. We compare fine-tuned LLMs against zero-shot LLMs and tabular classifiers (Figure~\ref{fig:experiment_overview}) with stratified 5-fold cross-validation, reporting means and 95\% confidence intervals. We evaluate individual-level classification with macro-F1 and distributional accuracy with aggregated vote shares and total variation distance (TVD).

\subsection{Survey Data Processing} \label{sec:method_data}

We use the German Longitudinal Election Study (GLES) Cross-Section Cumulation 2009--2017 \citep{glesGLESCrossSection200920172020}, which contains pre- and post-election face-to-face interviews. We predict vote choices in the 2017 post-election cross-section, enabling comparison with \citet{heydeVoxPopuliVox2024}.

\paragraph{Item selection and mapping.}
Survey items were selected based on factors associated with vote choice in German elections \citep{schmitt-beckNewEraElectoral2022,schoenVotersVotingContext2017,kleinGesellschaftlicheWertorientierungenWertewandel2014}: demographics (age, gender, education, income, employment status), religiosity, left--right ideology, party identification (and its strength), place of residence, and attitudes toward immigration and income inequality. Each item maps one-to-one to a feature, and the participant's self-reported vote choice is the target. We drop the infrequent categories \textit{invalid vote} and \textit{don't know}, and keep the original German wording since translation could obscure cultural nuance.

\paragraph{Missing values.}
Following \citet{heydeVoxPopuliVox2024}, we drop respondents with missing vote choices. However, we \emph{retain} other missing values rather than imputing them, since the type of missingness (e.g., \textit{no answer}, \textit{interview aborted}, or higher missingness for \textit{income}) may itself be informative.

\subsection{LLM Setup}\label{sec:method_llm}

\paragraph{Prompt design.}
Our task is instruction-based, so we use instruction-tuned chat models. Each instruction consists of a system prompt defining the task and a user prompt filled with a participant's survey responses (Figure~\ref{fig:experiment_overview}); the assistant response is the predicted vote choice. During fine-tuning we supervise on the assistant response; at evaluation only the system and user prompts are presented.

\paragraph{Model selection.}
For robustness, we test a range of model sizes across two families.
We use three Llama-3 instruction-tuned sizes: Llama-3.2-1B and Llama-3.2-3B~\citep{metaLlama32EdgeMobile2024}, and Llama-3.1-8B~\citep{dubeyLlama3Herd2024} (we omit the ``Instruct'' suffix below). We additionally run Qwen2.5 models from 0.5B to 7B~\citep{qwenQwen25TechnicalReport2025} in Appendix~\ref{appx:qwen_performance}.

\paragraph{Supervised fine-tuning.}
We fine-tune on a 40\,GB A100 partition in 16-bit precision using LoRA \citep{huLoRALowRankAdaptation2022} with rank $r=256$ applied to all linear layers. We use rank-stabilised scaling ($\alpha/\sqrt{r}$ instead of $\alpha/r$, with scaling factor $\alpha=8$) \citep{kalajdzievskiRankStabilizationScaling2023}, which was proposed precisely to keep learning effective at such high ranks. Training uses batch size 1, which fits the memory budget of the GPU partition, and runs for three epochs; all other hyperparameters remain at their library defaults. Training is conducted on completions only, where the loss is computed only on the assistant response tokens and not on the system and user prompt tokens. Formally, given a sequence of tokens $\mathbf{x} = (x_1, x_2, \ldots, x_n)$ where tokens $x_1$ to $x_k$ represent the prompt and tokens $x_{k+1}$ to $x_n$ represent the completion, the loss function is:
$$\mathcal{L} = -\sum_{i=k+1}^{n} \log P(x_i | x_1, \ldots, x_{i-1})$$
This ensures that the model learns to generate appropriate completions without being penalised for the fixed prompt content.\footnote{\url{https://huggingface.co/docs/trl/v0.13.0/en/sft_trainer}}

\paragraph{Answer extraction and probabilities.}
We follow \citet{argyleOutOneMany2023} and take the highest-probability token at the first answer position; the eight vote-choice categories (CDU/CSU, SPD, Greens, FDP, Die Linke, AfD, small party, non-voter) each have a distinct initial token, so this is unambiguous. We obtain a probability distribution by applying softmax only over the logits of these eight initial tokens. First-token probabilities are known to be unreliable in zero-shot settings: they need not match the model's actual text answer \citep{wangMyAnswerFirstToken2024}, and the elicited answer distributions are confounded by ordering and labelling biases \citep{dominguez-olmedoQuestioningSurveyResponses2023}. After fine-tuning, however, the mass on tokens outside the eight answer categories becomes negligible.

\subsection{Experiments}\label{sec:method_rqs}

All experiments use the same stratified 5-fold cross-validation splits, where each fold holds out 20\% of all GLES respondents as the test set. The training set is drawn from the remaining 80\% but varies by experiment: \textbf{E1} uses the full 80\%, simulating the scenario where responses are missing completely at random (RQ1). \textbf{E2--E4} restrict the training set to biased convenience samples\footnote{We use ``convenience sample'' as shorthand for biased subsamples constructed from the GLES data to mimic convenience sampling. These
simulate the effect of training on systematically restricted respondent groups, but are not \textit{real} convenience samples (e.g.,
respondents self-selecting into a non-probability survey).} within that 80\%, simulating systematic missingness (RQ2):
\begin{enumerate}
   \item[] \textbf{E2:} We restrict the training set to GLES participants whose reported occupation is ``student'' (enrolled in school or university), which amounts to 6\% of all participants. This simulates a scenario in which a researcher can only recruit students—a common convenience sample in the social sciences \citep{mullinixGeneralizabilitySurveyExperiments2015}—and must generalise from this subpopulation to the full electorate.
   \item[] \textbf{E3:} We restrict the training set to GLES participants residing in the state of Thuringia (5\% of participants). This simulates a regionally limited survey. As a former East German state, Thuringia had a distinctly different 2017 voting profile from the national average; for example, AfD's second-vote share was substantially higher in Thuringia than nationally~\citep{bundeswahlleiterin_btw2017_thuringia_results}. This makes it a useful regionally biased training subsample.
   \item[] \textbf{E4:} We restrict the training set to GLES participants who reported being unemployed (2\% of participants). This is the smallest and most biased convenience sample we test, motivated by evidence that unemployed individuals are more likely to participate in surveys when monetary incentives are offered \citep{singer2002paying}.
\end{enumerate}

These experiments pose a domain generalisation problem: restricting training to a specific subpopulation changes not only the covariate distribution $P_{\text{train}}(X) \neq P_{\text{test}}(X)$, but likely also the conditional $P(Y \mid X)$, since subpopulation membership may correlate with factors not included in our feature set (e.g., social milieu, peer effects) that also influence vote choice.

We additionally run each experiment in two variants: variant \textbf{a} uses all survey variables, while variant \textbf{b} removes ``party identification'', which is by far the strongest predictor of vote choice in our feature set (see Appendix~\ref{appx:party_identification_impact}) and tends to be missing whenever vote choice is also missing. This makes the ``b'' variants more realistic for practical settings.

\paragraph{Baselines.}
We compare against three \textbf{tabular classifiers} (logistic regression, random forest, CatBoost), \textbf{zero-shot LLMs} (GPT-4o, 2024-11-20, and our open-source models without fine-tuning), and a \textbf{prior-matched random classifier} that samples from the training label distribution. We also evaluate \textbf{importance-weighted variants} of the tabular classifiers to correct for covariate shift in the convenience sample experiments (Table~\ref{appx:tab_iw}).

\paragraph{Evaluation metrics.}
We are interested in assessing the predictive performance for individual respondents, as well as the aggregated vote share of the population. We use the following metrics:
\begin{itemize}
  \item \textbf{Macro-F1 score}: The macro-F1 score of the predicted vote choices, calculated as $\text{F1}_{\text{macro}} = \frac{1}{k} \sum_{i=1}^{k} \text{F1}_i$, where $\text{F1}_i$ is the F1 score for party $i$ and $k$ is the number of vote choice categories.
  \item \textbf{Aggregated vote share}: We calculate the aggregated vote share of the predicted vote choices as $\hat{P}(i) = \frac{1}{N} \sum_{j=1}^{N} \hat{p}_j(i)$, where $N$ is the number of test set participants and $\hat{p}_j(i)$ is the predicted probability that participant $j$ votes for party $i$. The true aggregated vote share is $P(i) = \frac{1}{N} \sum_{j=1}^{N} y_j(i)$, where $y_j(i)$ is 1 if participant $j$ voted for party $i$ and 0 otherwise.
  \item \textbf{Total variation distance (TVD)}: $\text{TVD}(\hat{P}, P) = \frac{1}{2} \sum_{i=1}^{k} |\hat{P}(i) - P(i)|$, where $\hat{P}$ and $P$ are the predicted and true aggregated vote shares over the $k$ vote choice categories.
\end{itemize}

\paragraph{Reproducibility.}
All experiments use fixed random seeds. GLES Cross-Section Cumulation 2009--2017 is openly available \citep{glesGLESCrossSection200920172020}, and our code is released.\footnote{\url{https://github.com/tobihol/convenience-sample-finetuning}}

\section{Results}

\begin{figure}
    \centering
    \includegraphics[width=0.95\columnwidth]{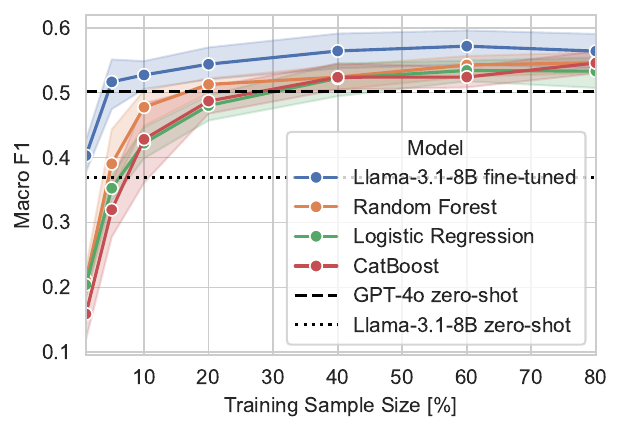}
    \caption{\textbf{Sample efficiency comparison.} We vary the size of the training sample (1\%, 5\%, 10\%, 20\%, 40\%, 60\%) of the E1a setup, drawn as \emph{random} sub-samples of GLES participants. Additionally, the original 80\% training sample is included. Because the sweep uses random sub-samples (not convenience-style biased ones), it isolates sample-size effects from sampling-bias effects.}
    \label{fig:sample_efficiency}
\end{figure}

\begin{figure}
  \centering
  \includegraphics[width=0.99\columnwidth]{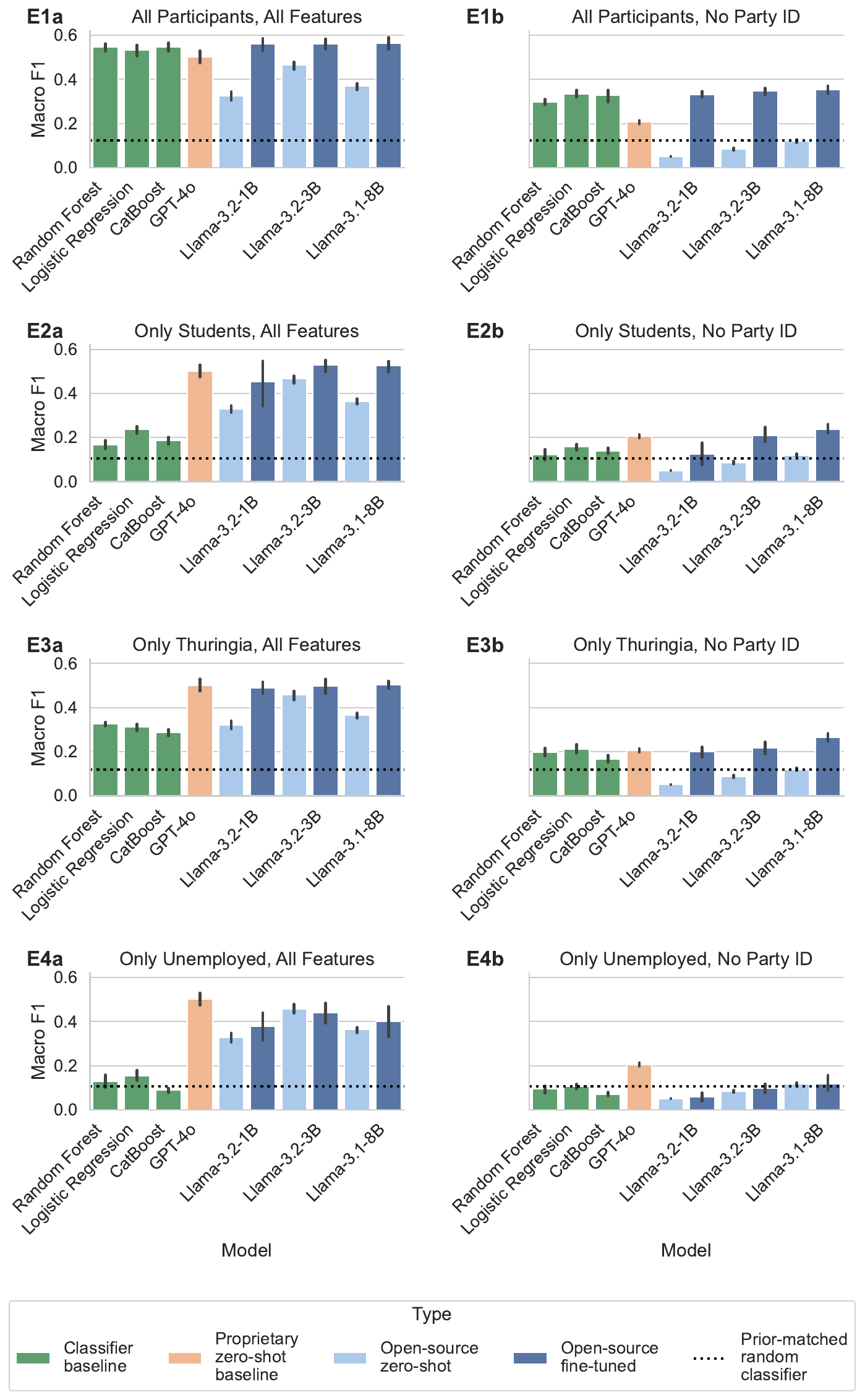}
  \caption{\textbf{Macro-F1 scores (higher is better)} of the predicted vote choices. The left column shows results using all survey variables as features. The right column shows results without using ``party identification''. The rows show different subsets of participants used for training.
  }
  \label{fig:combined_macro_f1}
\end{figure}

\subsection{RQ1: Missing Completely At Random}

\begin{figure}[!t]
  \centering
  \includegraphics[width=0.99\columnwidth]{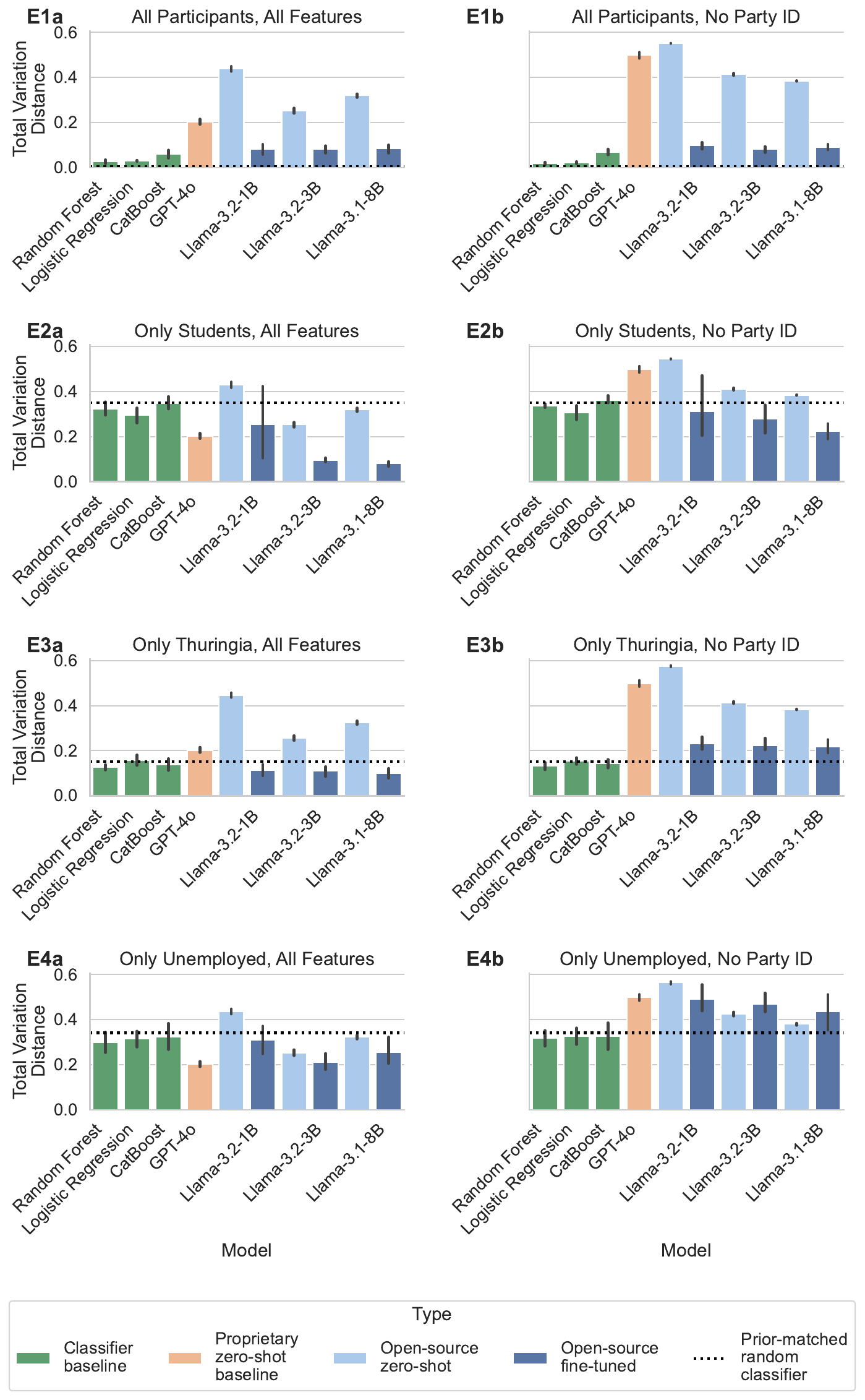}
  \caption{\textbf{Total variation distance (lower is better)} between the predicted distribution and the true distribution of vote choices. The left column shows results using all survey variables as features. The right column shows results without using ``party identification''. The rows show different subsets of participants used for training.
  }
  \label{fig:combined_total_variation}
\end{figure}

\begin{figure*}[!t]
    \centering
    \includegraphics[width=0.75\textwidth]{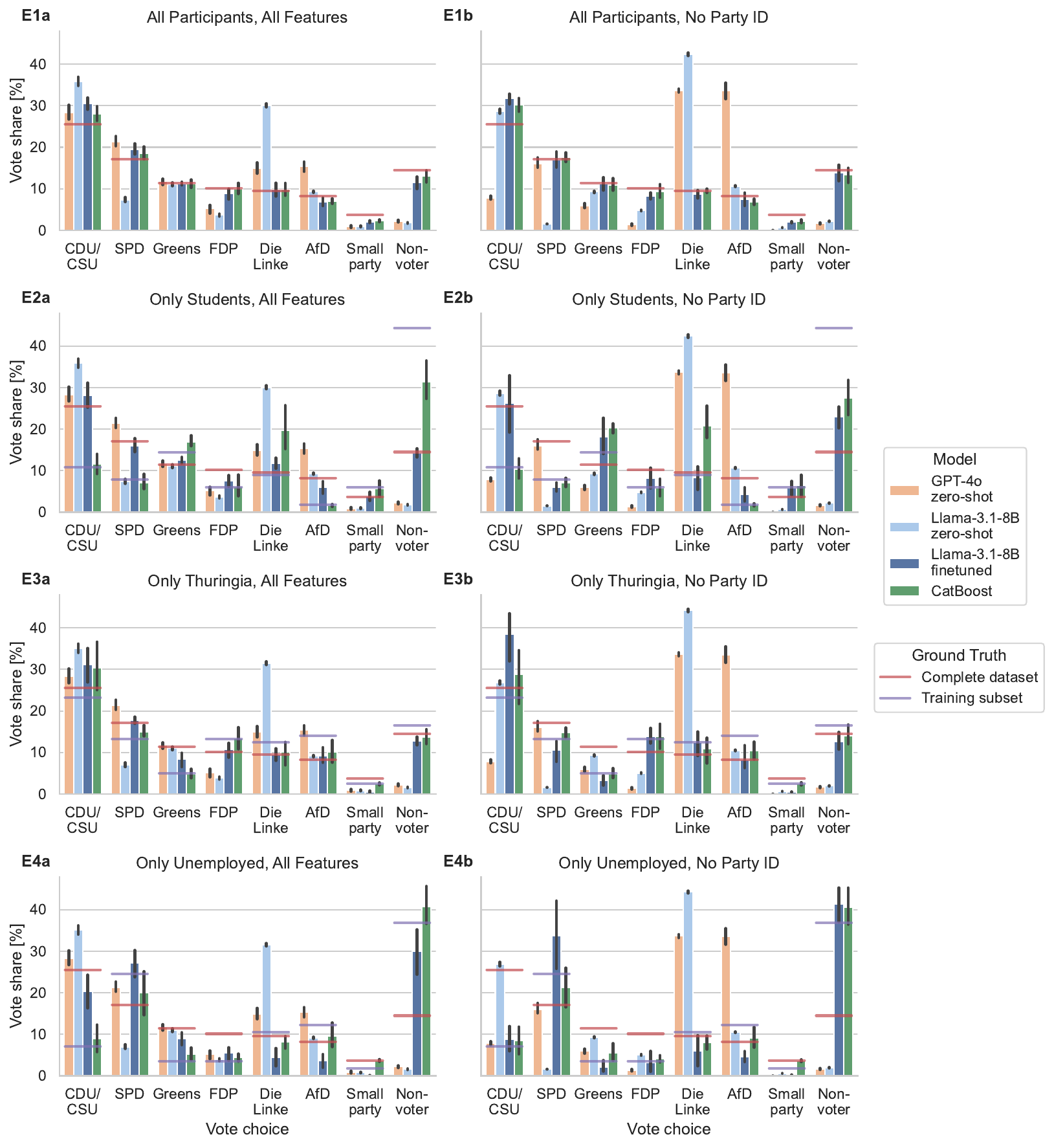}
    \caption{
    \textbf{Aggregated vote share} of the predicted vote choices by model across all experiments (E1--E4). The red line shows the true vote share in the complete dataset; the purple line shows the vote share within the respective biased training subset.
    For E1 both the tabular classifier and the fine-tuned model perform similarly well, in the case with and without having the strong predictor party identification in the data.
    The zero-shot models struggle with predicting non-voters in (a) and additionally overpredict more extreme parties (the AfD being the most right-wing party and Die Linke the most left-wing) in (b).
    For E2a, we see that fine-tuned LLMs are able to match the true distribution well. This degrades somewhat for E2b, and similarly to the CatBoost baseline, overfits to the student population by predicting more left-leaning parties and more non-voters.
    For E3, the training vote share distribution is more similar to the complete data. For E4, non-voters are again overrepresented in the training data, as in E2; here, however, the fine-tuned LLM does not generalise well to the overall population and overpredicts non-voters, similar to CatBoost.
    }
    \label{fig:combined_vote_share}
\end{figure*}

With a 20\% random test split, tabular classifiers and fine-tuned Llamas perform on par and both clearly outperform all zero-shot models in macro-F1, including GPT-4o (Figure~\ref{fig:combined_macro_f1}, E1). Zero-shot rankings are also unstable: e.g., Llama-3.1-8B trails the smaller Llama-3.2-3B in E1a, whereas the fine-tuned models are consistent across sizes and families. The same picture holds for TVD (Figure~\ref{fig:combined_total_variation}): fine-tuned and tabular models approximate the true distribution, while zero-shot models systematically deviate. Figure~\ref{fig:combined_vote_share} (E1) shows GPT-4o over-predicting parties at the extremes and rarely predicting non-voters, and zero-shot Llama-3.1-8B heavily over-predicting Die Linke. These patterns are consistent with the demographic and opinion-alignment biases documented by \citet{santurkarWhoseOpinionsLanguage2023} and with refusal behaviour, where the model avoids committing to certain answer categories \citep{wangMyAnswerFirstToken2024}.

The similar performance of all tabular classifiers and fine-tuned models in E1 suggests that when large random training samples are available, fine-tuning LLMs offers no clear advantage over traditional classification methods, which require less compute. The sample-size sweep in Figure~\ref{fig:sample_efficiency} refines this picture, however: zero-shot LLMs are only attractive below 5\% training data; from roughly 5\% up to about 40\% the fine-tuned Llama clearly outperforms all tabular classifiers; only as the sample approaches the full 80\% do tabular methods catch up. Because the sweep varies sample size on random sub-samples (not convenience-style biased ones), it isolates sample-size effects from sampling-bias effects. Removing the strong predictor ``party identification'' (E1a$\to$E1b) degrades all methods, but hurts zero-shot LLMs most (see gap between panels E1a and E1b in Figures~\ref{fig:combined_macro_f1} and \ref{fig:combined_total_variation}).

\subsection{RQ2: Convenience Samples}
As expected, the performance of all models drops when training data are limited to biased subsets of the GLES data. However, the extent of the decrease differs across experiments. In experiments E2 and E3, fine-tuned LLMs win: the 3B and 8B fine-tuned Llamas are better at modelling the true vote share distribution than both the zero-shot LLMs and the tabular classifiers. For the smallest sample (E4, 2\% of participants), the picture flips: zero-shot GPT-4o becomes the strongest model, and fine-tuned LLMs no longer outperform the tabular baselines. Applying importance weighting to correct for covariate shift in the tabular baselines yields mixed results: CatBoost benefits in some experiments, but the overall gap to fine-tuned LLMs remains (see Table~\ref{appx:tab_iw}).

\paragraph{E2: Only Student Data}
Figure~\ref{fig:combined_macro_f1} (E2a) shows that fine-tuned open-source LLMs and GPT-4o are better able to predict the vote choice of GLES participants in the test split, compared to tabular classifiers. Figure~\ref{fig:combined_total_variation} (E2a) indicates that the 3B and 8B fine-tuned Llama models in particular achieve considerably lower total variation distance, suggesting that fine-tuning can help to reduce the deviation from the true distribution, even when training on biased data.
Deviations, however, still exist. GPT-4o produces especially large deviations for non-voters and the extreme parties (Die Linke and AfD; see Figure~\ref{fig:combined_vote_share}). CatBoost also overpredicts non-voters and underpredicts CDU/CSU considerably.

When the important predictor party identification is removed from the training data, zero-shot models in particular start to overestimate the share of Die Linke and the AfD (cf. Figure~\ref{fig:combined_vote_share} E2b).

\paragraph{E3: Only Thuringia Region}
Figure~\ref{fig:combined_macro_f1} shows that F1 scores for E3 are higher across all models compared to E2, which is expected since the Thuringia training distribution is closer to the national vote distribution than the student sample (see Figure~\ref{fig:combined_vote_share}, training subset). In both variants, E3a and E3b, the 8B fine-tuned Llama model outperforms the classifier baseline with respect to the macro-F1. However, the difference in E3b is notably smaller than in E3a. Both fine-tuned LLMs and tabular classifiers decrease in performance when an important predictor variable is removed and show comparably poor performance.

Figure~\ref{fig:combined_total_variation} shows that the fine-tuned LLMs and tabular classifier perform similarly with respect to the TVD.
Figure~\ref{fig:combined_vote_share} also shows that the fine-tuned LLMs and tabular classifier make similar mistakes (e.g., they overestimate CDU/CSU and underestimate the Greens), directly reflecting Thuringia's training distribution, which over-represents CDU/CSU and AfD relative to the national average. Zero-shot LLMs again strongly overestimate the share of Die Linke and the AfD, as in the previous experiment.

\paragraph{E4: Only Unemployed Data}
This experiment shows the lowest scores for the fitted models, which is expected given that it uses the smallest training set. The fine-tuned models fail to outperform the zero-shot setting, as reflected by their similar F1 scores and total variation distances. In this case, the zero-shot proprietary model achieves the best performance among all models, but only marginally so. In Figure~\ref{fig:combined_vote_share}, we can see that both fine-tuned models and CatBoost overpredict non-voters for E4, which fine-tuned models did not do for E2, although the two training sets both over-represent non-voters relative to the complete dataset distribution (see the training-subset lines in Figure~\ref{fig:combined_vote_share}).

This experiment shows that for extremely small convenience samples (2\%), fine-tuned LLMs and tabular classifiers perform worse than zero-shot LLMs.

\section{Discussion and Conclusion}\label{sec:discussion}

The findings of this study contribute to the growing literature on the use of large language models for predicting survey responses. Our results show that the effectiveness of LLMs for predicting survey responses depends critically on the amount and composition of available training data.

When only small but biased convenience samples are available, such as datasets composed exclusively of students, fine-tuning small open-source LLMs (3B--8B) yields substantial gains: fine-tuned models recover both individual-level predictions and aggregate vote-share distributions more accurately than zero-shot LLMs, including GPT-4o, and often outperform tabular baselines. These gains likely arise because fine-tuned models can draw simultaneously on sample-specific correlations and broader knowledge that models have acquired in the pre-training phase.

However, the advantages of fine-tuning diminish when training sets become extremely small. In the E4 condition, where only 2\% of participants were available for training, fine-tuned models no longer outperformed their zero-shot counterparts, and the zero-shot proprietary model achieved the strongest results. This suggests that very small samples provide insufficient task-specific signal for effective supervised adaptation, making zero-shot inference from large pretrained models a more reliable choice.

When medium- to large-sized random samples are available (E1), fine-tuned LLMs and tabular classifiers perform similarly, and given the substantially higher computational cost of LLMs, we do not consider them promising for imputing small amounts of randomly missing data in surveys. A caveat is that zero-shot proprietary models like GPT-4o are not fully reproducible due to model updates and nondeterminism.

Our zero-shot results also speak to the concern of data contamination (see Limitations): if the models had simply memorised the widely documented 2017 election results, their zero-shot predictions should closely match the true aggregate vote shares. Instead, we observe substantial deviations in the zero-shot setting (e.g., systematic over-prediction of AfD and Die Linke, under-prediction of non-voters; cf.\ Figure~\ref{fig:combined_vote_share}), which provides indirect evidence against strong contamination.

It is important to note that even the best-performing models often achieve macro-F1 below 0.6 in the convenience-sample settings (E2--E4). Our results should therefore be read as evidence that fine-tuned LLMs are a \textit{promising direction} rather than a ready-to-deploy solution, especially in multi-party systems like Germany's where small parties can decisively influence coalitions.

Taken together, these findings offer practical guidance, although we caution that they are derived from a single election in a single country and from convenience samples constructed retrospectively (see Limitations): when training data are extremely limited, zero-shot prompting of a strong proprietary model remained competitive in our experiments. When small-to-medium but biased convenience samples are available, fine-tuning small open-source LLMs was the most promising approach in our setting, as these models leverage both sample-specific signals and world knowledge. With medium-to-large random samples that reflect the target population, traditional machine-learning models remain strong options.

Future work should test generalisation to other survey variables, populations, and response formats, and systematically establish the amount and type of training data required for fine-tuning to yield stable improvements.

\section*{Limitations}\label{sec:limitations}

A potential concern is data contamination: the 2017 German federal election results are widely documented online and likely part of the pre-training corpora of the models we evaluate (Llama-3, GPT-4o). In principle, LLMs could leverage memorised aggregate statistics rather than reasoning about individual respondent profiles. While the zero-shot deviations discussed in Section~\ref{sec:discussion} argue against strong contamination, we cannot fully rule out indirect advantages from pre-training exposure, and future work should evaluate on elections that post-date the models' training cutoff.

Our convenience samples are also constructed retrospectively from a probability-based survey rather than collected as real non-probability samples. This design lets us compare models against a common held-out test set and isolate the effect of biased training composition, but it may understate or overstate the difficulties of real convenience samples, where recruitment, motivation, survey mode, and measurement error can differ from probability samples.

Temporal validity poses a further concern: voting behaviour shifts across election cycles, and it is unclear how our findings transfer to later elections (e.g., 2021 or 2025).
We also use zero-shot LLMs as baselines without prompt variation, acknowledging that their performance could change with alternative prompts. Finally, while we focused on fine-tuning, techniques like retrieval-augmented generation or in-context learning with long-context models are promising directions for future work.

\section*{Ethical Considerations}\label{sec:ethics}
The data used in this study are derived from survey results available through GESIS under access conditions, with all responses anonymised. The survey data contain no personally identifying information or offensive content.
An important ethical concern with simulating public opinions using LLMs is that generated responses may misrepresent genuine public sentiment. Because LLMs base their outputs on patterns found in historical and potentially biased data, they could reinforce prejudices, obscure minority perspectives, or amplify simplistic narratives. Our results suggest that fine-tuning can help reduce but not fully mitigate these issues.

Beyond misrepresentation, predicting voting behaviour itself carries risks for democratic processes. Models that infer vote choice from socio-demographic and attitudinal profiles could be misused for targeted political advertising or demobilisation campaigns aimed at groups predicted to vote a certain way, and synthetic ``respondents'' could be passed off as genuine public opinion to influence public debate or election coverage. Inaccurate predictions are also a danger in their own right if they are treated as reliable, for example when published as election forecasts. This is one reason we report absolute performance levels prominently and characterise our results as methodological rather than as forecasts. We therefore see the methods studied here as tools for research on inference from non-probability samples, not as substitutes for representative surveys. Moreover, vote choice is protected by the secrecy of the ballot: predictions of individual vote choice must not be used to profile or target identifiable individuals. Our study works exclusively with anonymised, retrospective academic survey data used under the GLES usage agreement, and we evaluate only aggregate and anonymised individual-level predictive performance.

\section*{Acknowledgements}
This work was supported by the German Research Foundation (DFG, project no.~504226141) and by the DAAD programme Konrad Zuse Schools of Excellence in Artificial Intelligence, sponsored by the Federal Ministry of Research, Technology and Space.

\bibliography{bibliography}

\clearpage
\appendix

\section{Importance Weighting of Classifier Baselines}\label{appx:importance_weighting}

To assess whether the tabular classifier baselines in the convenience sample experiments (E2--E4) can be improved by correcting for covariate shift, we apply importance weighting \citep{shimodaira2000improving}. The density ratio $w(x) = p_{\text{test}}(x)\,/\,p_{\text{train}}(x)$ is estimated directly using RuLSIF (Relative unconstrained Least-Squares Importance Fitting) \citep{yamadaRelativeDensityratioEstimation2011}, a kernel-based method that avoids separate density estimation for numerator and denominator. Training samples are then reweighted by the estimated density ratio and passed to the classifier via the \texttt{sample\_weight} parameter.

\begin{table}[ht]
    \centering
    \resizebox{\columnwidth}{!}{\begin{tabular}{llcccc}
\toprule
 &  & \multicolumn{2}{c}{Macro F1} & \multicolumn{2}{c}{TVD} \\
 &  & Vanilla & IW & Vanilla & IW \\
Exp. & Model &  &  &  &  \\
\midrule
\multirow[t]{3}{*}{E2a} & LogReg & $0.237 \pm 0.019$ & $0.219 \pm 0.018$ & $0.295 \pm 0.036$ & $0.282 \pm 0.036$ \\
 & RF & $0.169 \pm 0.022$ & $0.167 \pm 0.026$ & $0.322 \pm 0.032$ & $0.328 \pm 0.035$ \\
 & CatBoost & $0.186 \pm 0.019$ & $0.199 \pm 0.038$ & $0.349 \pm 0.031$ & $0.407 \pm 0.068$ \\
\cline{1-6}
\multirow[t]{3}{*}{E2b} & LogReg & $0.158 \pm 0.015$ & $0.161 \pm 0.013$ & $0.306 \pm 0.034$ & $0.294 \pm 0.036$ \\
 & RF & $0.122 \pm 0.027$ & $0.109 \pm 0.013$ & $0.337 \pm 0.009$ & $0.333 \pm 0.019$ \\
 & CatBoost & $0.139 \pm 0.015$ & $0.135 \pm 0.020$ & $0.362 \pm 0.020$ & $0.427 \pm 0.047$ \\
\cline{1-6}
\multirow[t]{3}{*}{E3a} & LogReg & $0.311 \pm 0.017$ & $0.304 \pm 0.012$ & $0.156 \pm 0.026$ & $0.153 \pm 0.028$ \\
 & RF & $0.325 \pm 0.010$ & $0.330 \pm 0.010$ & $0.126 \pm 0.013$ & $0.123 \pm 0.015$ \\
 & CatBoost & $0.286 \pm 0.016$ & $0.327 \pm 0.015$ & $0.138 \pm 0.030$ & $0.160 \pm 0.019$ \\
\cline{1-6}
\multirow[t]{3}{*}{E3b} & LogReg & $0.212 \pm 0.022$ & $0.202 \pm 0.025$ & $0.154 \pm 0.017$ & $0.144 \pm 0.020$ \\
 & RF & $0.199 \pm 0.020$ & $0.180 \pm 0.016$ & $0.132 \pm 0.018$ & $0.138 \pm 0.016$ \\
 & CatBoost & $0.167 \pm 0.018$ & $0.211 \pm 0.013$ & $0.144 \pm 0.021$ & $0.167 \pm 0.012$ \\
\cline{1-6}
\multirow[t]{3}{*}{E4a} & LogReg & $0.154 \pm 0.027$ & $0.136 \pm 0.028$ & $0.315 \pm 0.040$ & $0.278 \pm 0.047$ \\
 & RF & $0.130 \pm 0.033$ & $0.133 \pm 0.029$ & $0.299 \pm 0.051$ & $0.302 \pm 0.047$ \\
 & CatBoost & $0.089 \pm 0.011$ & $0.165 \pm 0.031$ & $0.325 \pm 0.065$ & $0.333 \pm 0.062$ \\
\cline{1-6}
\multirow[t]{3}{*}{E4b} & LogReg & $0.107 \pm 0.012$ & $0.109 \pm 0.016$ & $0.326 \pm 0.041$ & $0.301 \pm 0.043$ \\
 & RF & $0.094 \pm 0.018$ & $0.094 \pm 0.013$ & $0.317 \pm 0.039$ & $0.323 \pm 0.048$ \\
 & CatBoost & $0.070 \pm 0.010$ & $0.105 \pm 0.008$ & $0.327 \pm 0.066$ & $0.357 \pm 0.055$ \\
\bottomrule
\end{tabular}
}
    \caption{Macro-F1 and TVD ($\text{mean} \pm \text{95\% CI}$) for vanilla and importance-weighted (IW) classifier baselines across convenience sample experiments E2--E4.}
    \label{appx:tab_iw}
  \end{table}

Table~\ref{appx:tab_iw} shows that importance weighting has mixed effects across models. For logistic regression and random forest, macro-F1 and TVD remain largely unchanged, with overlapping confidence intervals. CatBoost, however, shows notable macro-F1 improvements with importance weighting in E3 and E4 (e.g., E4a: $0.089 \to 0.165$), likely because the reweighting helps its more flexible decision boundaries generalise better. At the same time, CatBoost's TVD increases in E2, indicating that improved per-sample accuracy does not necessarily translate to better aggregate vote share estimation. Overall, even where importance weighting helps, the resulting performance remains well below that of fine-tuned LLMs, suggesting that the gap is not primarily caused by covariate shift in the feature space.

\setlength{\floatsep}{6pt plus 2pt minus 2pt}
\setlength{\textfloatsep}{8pt plus 2pt minus 2pt}
\setlength{\abovecaptionskip}{4pt}

\vfill\newpage

\section{Robustness Across Model Families}\label{appx:qwen_performance}

\begin{figure}[!htb]
    \centering
    \includegraphics[width=0.78\columnwidth]{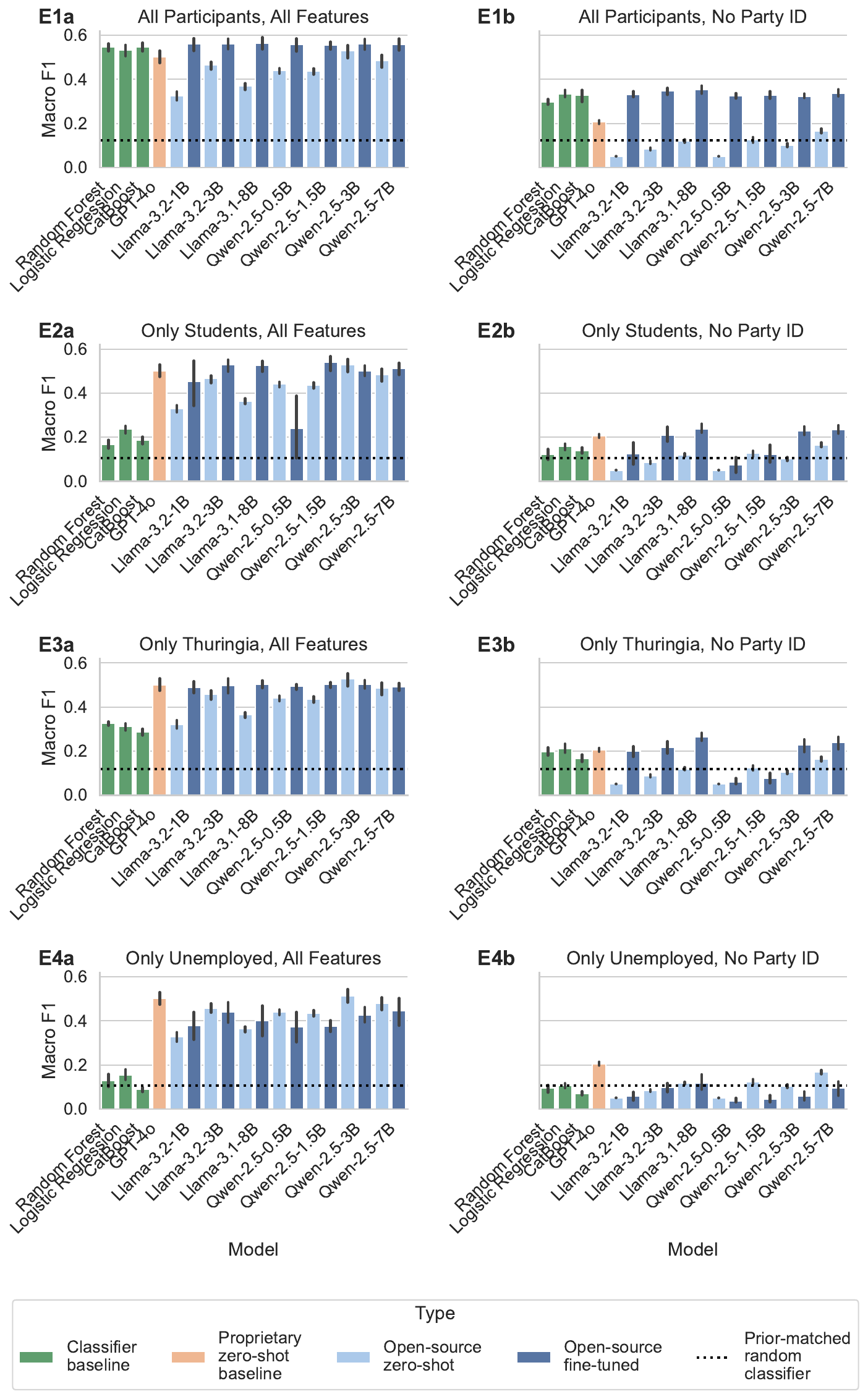}
    \caption{Macro-F1 scores of Llama and Qwen models up to 8B parameters.}
    \label{appx:rq1_performance_by_model_macro_f1}
\end{figure}

\begin{figure}[!htb]
    \centering
    \includegraphics[width=0.78\columnwidth]{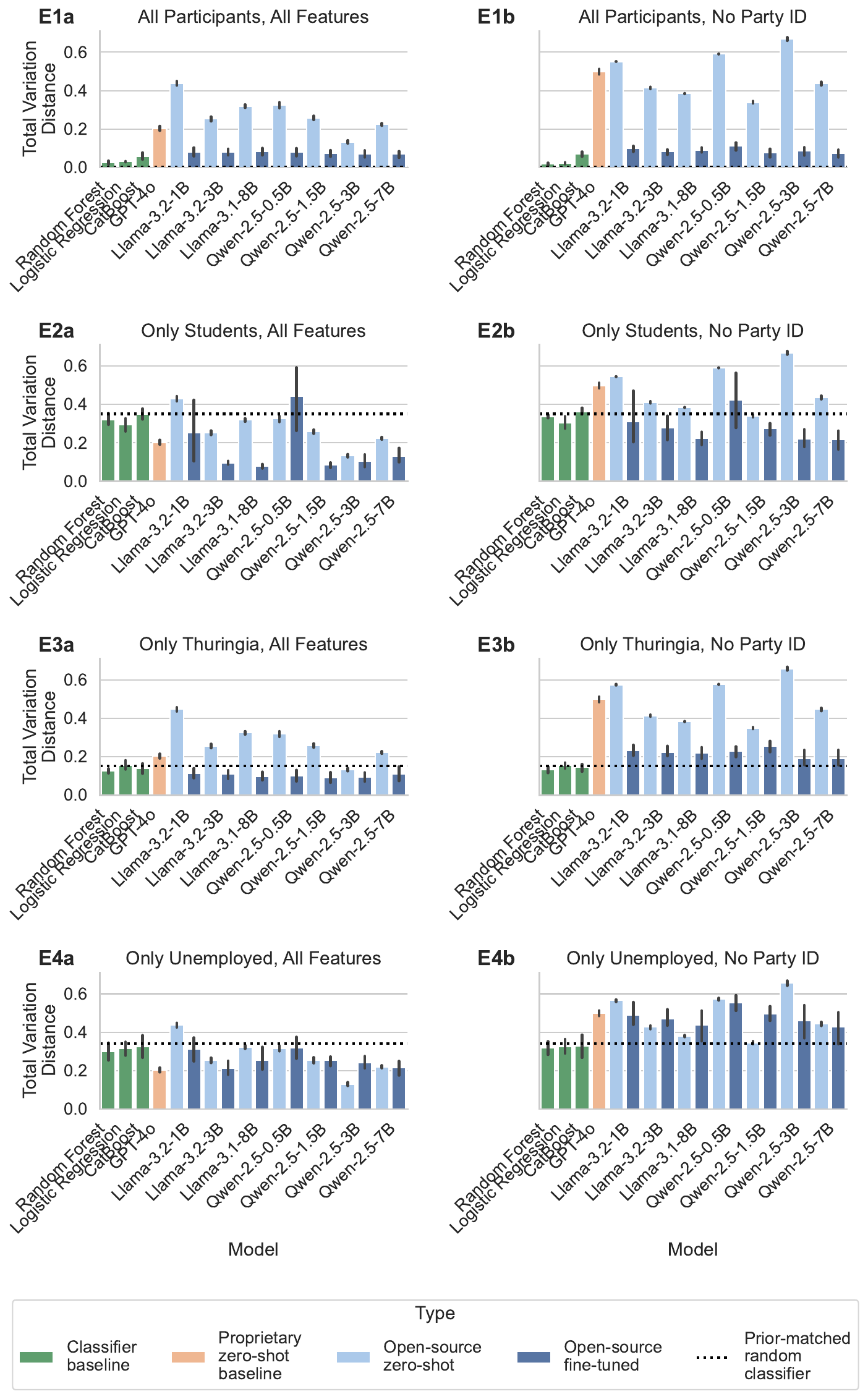}
    \caption{TVD of Llama and Qwen models up to 8B parameters.}
    \label{appx:rq1_performance_by_model_tvd}
\end{figure}

\clearpage

\section{Party Identification Impact}\label{appx:party_identification_impact}

\begin{figure}[ht]
    \centering
    \includegraphics[width=\columnwidth]{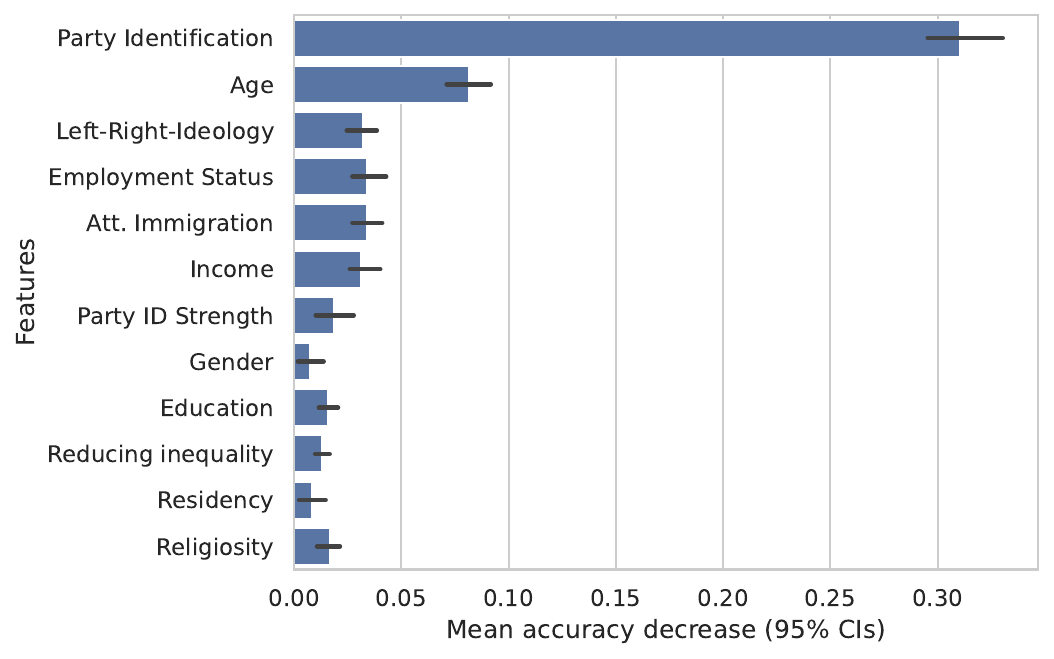}
    \caption{Feature importance for the logistic regression classifier in E1a computed using feature permutation, where each feature's values are randomly shuffled to quantify the drop in performance. This drop indicates how much each feature contributes to the model's predictive power.}
    \label{appx_fig:lr_feature_importance}
\end{figure}

\begin{figure}[ht]
    \centering
    \includegraphics[width=\columnwidth]{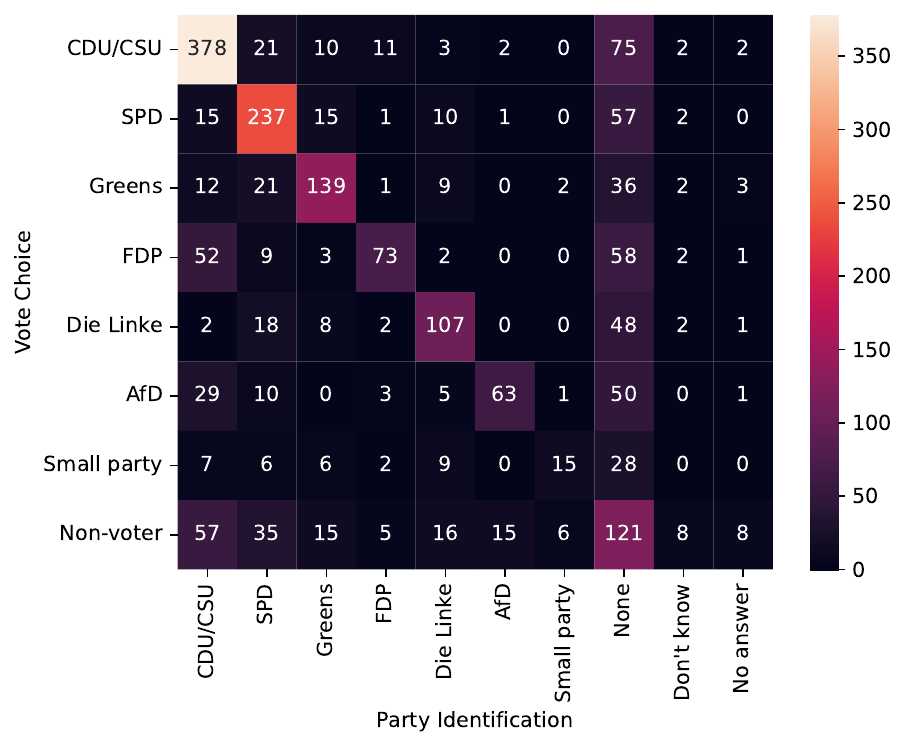}
    \caption{Cross-tabulation of ``vote choice'' and ``party identification''.}
    \label{appx_fig:vote_choice_and_party_id_cross_tabulation}
\end{figure}

Figure~\ref{appx_fig:lr_feature_importance} shows permutation feature importance for the logistic regression baseline in E1a: shuffling ``party identification'' decreases accuracy by roughly 0.31, around four times the drop of the next most important feature (age). The cross-tabulation in Figure~\ref{appx_fig:vote_choice_and_party_id_cross_tabulation} illustrates why: most respondents report voting for the party they identify with. This dominant role motivates the ``b'' variants of our experiments, which remove party identification from the training features.

\vfill\newpage

\end{document}